\documentclass[11pt,a4paper,]{article}
\usepackage{lmodern}

\usepackage{amssymb,amsmath}
\usepackage{ifxetex,ifluatex}
\usepackage{fixltx2e} 
\ifnum 0\ifxetex 1\fi\ifluatex 1\fi=0 
  \usepackage[T1]{fontenc}
  \usepackage[utf8]{inputenc}
\else 
  \usepackage{unicode-math}
  \defaultfontfeatures{Ligatures=TeX,Scale=MatchLowercase}
\fi
\IfFileExists{upquote.sty}{\usepackage{upquote}}{}
\IfFileExists{microtype.sty}{%
\usepackage[]{microtype}
\UseMicrotypeSet[protrusion]{basicmath} 
}{}
\PassOptionsToPackage{hyphens}{url} 
\usepackage[unicode=true]{hyperref}
\hypersetup{
            pdftitle={COVID-19 and Online Learning Tools},
            pdfkeywords={Online learning, Online teaching Distance education solutions, COVID-19 Pandemic, Google Trend search queries},
            pdfborder={0 0 0},
            breaklinks=true}
\usepackage{geometry}
\usepackage[style=authoryear-comp,]{biblatex}
\usepackage{longtable,booktabs}
\IfFileExists{footnote.sty}{\usepackage{footnote}\makesavenoteenv{long table}}{}
\IfFileExists{parskip.sty}{%
\usepackage{parskip}
}{
\setlength{\parindent}{0pt}
\setlength{\parskip}{6pt plus 2pt minus 1pt}
}
\makeatletter
\def\fps@figure{htbp}
\makeatother

\title{COVID-19 and Online Learning Tools}

\RequirePackage{caption}
\DeclareCaptionStyle{italic}[justification=centering]
 {labelfont={bf},textfont={it},labelsep=colon}
\RequirePackage{bera}
\RequirePackage{mathpazo}

\RequirePackage{fancyhdr}
\fancypagestyle{plain}{%
\fancyhf{} 
\fancyfoot[C]{\sffamily\thepage} 

}

\RequirePackage{bm,amsmath}
\allowdisplaybreaks

\RequirePackage{graphicx}
\RequirePackage[compact,sf,bf]{titlesec}
\titleformat{\section}[block]
  {\fontsize{15}{17}\bfseries\sffamily}
  {\thesection}
  {0.4em}{}
\titleformat{\subsection}[block]
  {\fontsize{12}{14}\bfseries\sffamily}
  {\thesubsection}
  {0.4em}{}
\titlespacing{\section}{0pt}{*5}{*1}
\titlespacing{\subsection}{0pt}{*2}{*0.2}

\def\Date{\number\day}
\def\Month{\ifcase\month\or
 January\or February\or March\or April\or May\or June\or
 July\or August\or September\or October\or November\or December\fi}
\def\Year{\number\year}

\makeatletter
\def\wp#1{\gdef\@wp{#1}}\def\@wp{??/??}
\def\jel#1{\gdef\@jel{#1}}\def\@jel{??}
\def\showjel{{\large\textsf{\textbf{JEL classification:}}~\@jel}}

\def\addresses#1{\gdef\@addresses{#1}}\def\@addresses{??}
\def\cover{{\sffamily\setcounter{page}{0}
        \thispagestyle{empty}
        \vspace*{2cm}
        \begin{center}
        \fbox{\parbox{14cm}{\begin{onehalfspace}\centering\Huge\vspace*{0.3cm}
                \textsf{\textbf{\expandafter{\@title}}}\vspace{1cm}\par
                \LARGE\@author\end{onehalfspace}
        }}
        \end{center}
        \vfill
                \begin{center}\Large
                \Month~\Year\\[1cm]
                Working Paper \@wp
        \end{center}\vspace*{2cm}}}
\def\pageone{{\sffamily\setstretch{1}%
        \thispagestyle{empty}%
        \vbox to \textheight{%
        \raggedright\baselineskip=1.2cm
     {\fontsize{24.88}{30}\sffamily\textbf{\expandafter{\@title}}}
        \vspace{2cm}\par
        \hspace{1cm}\parbox{14cm}{\sffamily\large\@addresses}\vspace{1cm}\vfill
        \hspace{1cm}{\large\Date~\Month~\Year}\\[1cm]
        \hspace{1cm}\showjel\vss}}}
\def\blindtitle{{\sffamily
     \thispagestyle{plain}\raggedright\baselineskip=1.2cm
     {\fontsize{24.88}{30}\sffamily\textbf{\expandafter{\@title}}}\vspace{1cm}\par
        }}
\def\titlepage{{\cover\newpage\pageone\newpage\blindtitle}}

\def\blind{\def\titlepage{{\blindtitle}}\let\maketitle\blindtitle}
\def\titlepageonly{\def\titlepage{{\pageone\end{document}}}}
\def\nocover{\def\titlepage{{\pageone\newpage\blindtitle}}\let\maketitle\titlepage}
\let\maketitle\titlepage
\makeatother

\RequirePackage{setspace}
\spacing{1.5}

\RequirePackage{microtype}

\RequirePackage{xcolor} 
\definecolor{darkblue}{rgb}{0,0,.6}
\RequirePackage{url}

\makeatletter
\@ifpackageloaded{hyperref}{}{\RequirePackage{hyperref}}
\makeatother
\hypersetup{
     citecolor=0 0 0,
     breaklinks=true,
     bookmarksopen=true,
     bookmarksnumbered=true,
     linkcolor=darkblue,
     urlcolor=blue,
     citecolor=darkblue,
     colorlinks=true}

\newenvironment{keywords}{\par\vspace{0.5cm}\noindent{\sffamily\textbf{Keywords:}}}{\vspace{0.25cm}\par\hrule\vspace{0.5cm}\par}

\renewenvironment{abstract}{\begin{minipage}{\textwidth}\parskip=1.4ex\noindent
\hrule\vspace{0.1cm}\par{\sffamily\textbf{\abstractname}}\newline}
  {\end{minipage}}

\usepackage[T1]{fontenc}
\usepackage[utf8]{inputenc}

\usepackage[showonlyrefs]{mathtools}
\usepackage[no-weekday]{eukdate}

\makeatletter
\@ifpackageloaded{biblatex}{}{\usepackage[style=authoryear-comp, backend=biber, natbib=true]{biblatex}}
\makeatother
\DeclareFieldFormat{url}{\texttt{\url{#1}}}
\DeclareFieldFormat[article]{pages}{#1}
\DeclareFieldFormat[inproceedings]{pages}{\lowercase{pp.}#1}
\DeclareFieldFormat[incollection]{pages}{\lowercase{pp.}#1}
\DeclareFieldFormat[article]{volume}{\mkbibbold{#1}}
\DeclareFieldFormat[article]{number}{\mkbibparens{#1}}
\DeclareFieldFormat[article]{title}{\MakeCapital{#1}}
\DeclareFieldFormat[inproceedings]{title}{#1}
\DeclareFieldFormat{shorthandwidth}{#1}
\usepackage{xpatch}
\xpatchbibmacro{volume+number+eid}{\setunit*{\adddot}}{}{}{}
\renewbibmacro{in:}{%
  \ifentrytype{article}{}{%
  \printtext{\bibstring{in}\intitlepunct}}}

\makeatletter
\DeclareDelimFormat[cbx@textcite]{nameyeardelim}{\addspace}
\makeatother

\wp{1}
\jel{C82,I20,I21}

\RequirePackage[absolute,overlay]{textpos}
\author{Priyanga Dilini~Talagala, Thiyanga S.~Talagala}
\addresses{\textbf{Priyanga Dilini Talagala}\newline
Department of Compuational Mathematics, University of Moratuwa, Sri Lanka
\newline{Email: \href{mailto:priyangad@uom.lk}{\nolinkurl{priyangad@uom.lk}}}\newline Corresponding author\\[1cm]
\textbf{Thiyanga S. Talagala}\newline
Department of Statistics, Faculty of Applied Sciences, University of Sri Jayewardenepura, Sri Lanka
\newline{Email: \href{mailto:ttalagala@sjp.ac.lk}{\nolinkurl{ttalagala@sjp.ac.lk}}}\\[1cm]
}

\date{\sf\Date~\Month~\Year}

\begin{document}
\maketitle
\begin{abstract}
Distance education has a long history. However, COVID-19 has created a new era of distance education. Due to the increasing demand, various distance learning solutions have been introduced for different distance education purposes. In this study, we investigated the impact of COVID-19 on global attention towards different distance learning-teaching tools. We used Google Trend search queries as a proxy to quantify the popularity and public interest towards different distance education solutions. Both visual and analytical approaches were used to analyse global level web search queries during the COVID-19 pandemic. This can provide a fast first step guide to identifying the most popular online learning tools available for different educational purposes. The results allow the teachers to narrow down the search space and deepen their exploration of prominent distance education solutions to support their online teaching. The R code and data to reproduce the results of this work are available in the online supplementary materials.
\end{abstract}
\begin{keywords}
Online learning, Online teaching Distance education solutions, COVID-19 Pandemic, Google Trend search queries
\end{keywords}

\hypertarget{introduction}{%
\section{Introduction}\label{introduction}}

COVID-19 is still one of the most important priorities of governments and media in many countries all around the world. Due to the alarming levels of spread and severity of the virus, the World Health Organization (WHO) declared the outbreak a Public Health Emergency of International Concern (PHEIC) on January 30, 2020, and then a pandemic on March 11, 2020 \autocite{world2020timeline}. As of September 29, 2021, more than 232.7 million cases of COVID-19 have been reported in over 192 countries and territories, resulting in more than 4.7 million deaths \autocite{dong2020interactive}. In response to the pandemic, health authorities worldwide have taken many steps, including vaccination development and deployment to minimize the spread of the virus. In addition, authorities worldwide have also taken many non-pharmaceutical interventions and preventive measures such as travel restrictions, lock-downs, workplace hazard controls, school/university closures, facility closures, and reduction of mass gatherings to reduce the spread of the virus \autocite{chang2020modelling}.

\begin{figure}[h]

{\centering \includegraphics[width=1\textwidth]{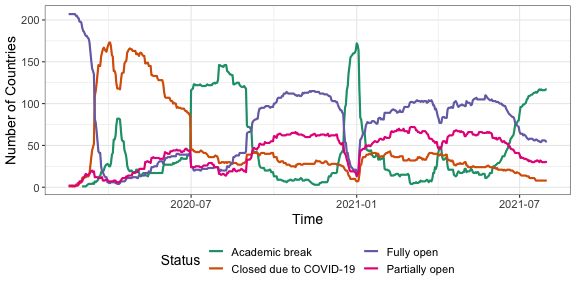} 

}

\caption{Global tracking of COVID-19 caused school closures and re-openings}\label{fig:covidImpactWorld}
\end{figure}

\begin{figure}[h]

{\centering \includegraphics[width=1\textwidth]{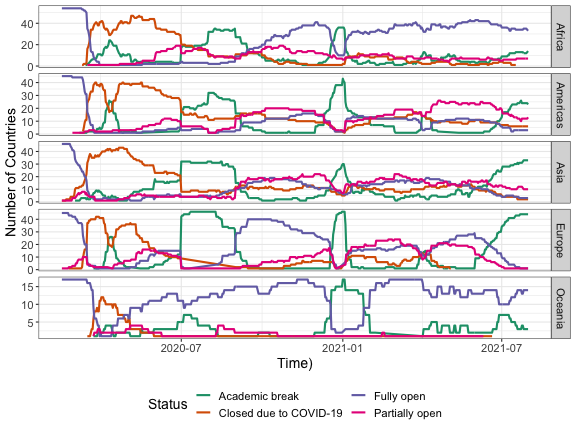} 

}

\caption{Region wise tracking of COVID-19 caused school closures and re-openings}\label{fig:covidImpactContinent}
\end{figure}

Many sectors have been affected ever since the outbreak of COVID-19, worldwide. Among these many sectors, education is one of the most affected sectors with a near-total closure of schools, colleges, and universities all around the world \autocite{daniel2020education}. Figures \ref{fig:covidImpactWorld} show the evolution of global school closures and reopenings since mid February 2020 \autocite{unesco2020covid}. With the start of the pandemic, a near total closure of schools was observed all around the world. Over time, Africa and Oceania demonstrated better recovery in comparison to other regions with their increasing number of fully open schools in which the classes are held exclusively in person (Figures \ref{fig:covidImpactContinent}).

\begin{figure}[h]

{\centering \includegraphics[width=1\textwidth]{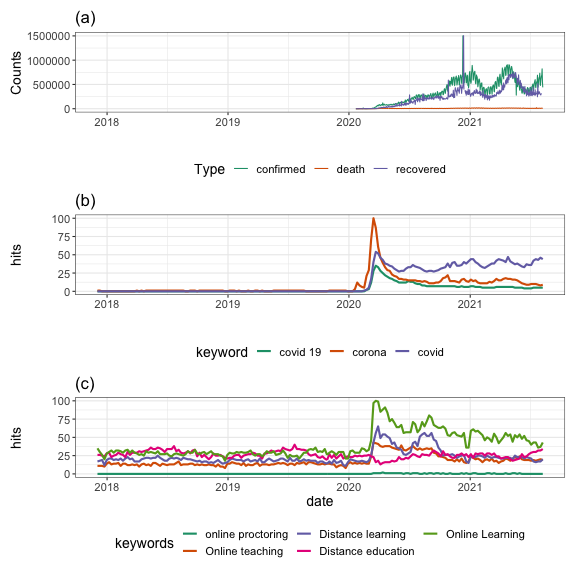} 

}

\caption{Visualization of data from Goolge search trends. (a) COVID-19 cases worldwide (b) Google search trends of COVID-19 related terms (c) Google search trends of distance education related terms.}\label{fig:distanceLearningWorldAnalysis}
\end{figure}

Even though distance education has a long history that goes back almost two centuries with significant modifications,
alterations, and additions in the process of delivery and communication \autocite{moore2011learning,spector2014handbook}, COVID-19 certainly created a new era of distance education while encouraging various stakeholders in education to take the concept of distance education seriously \autocite{richmond2020critical}. Sudden unexpected movement from classrooms to home schooling at large-scale made students, teachers and parents vulnerable, leading to millions of education related internet searches being performed during the period of COVID-19 pandemic (Figures \ref{fig:distanceLearningWorldAnalysis}) \autocite{carter2021teacher}. Surprisingly, the search spikes of distance education related searches (Figure \ref{fig:distanceLearningWorldAnalysis} (c)) coincide with increasing COVID-19 counts (Figure \ref{fig:distanceLearningWorldAnalysis} (a)) and related internet searches (Figure \ref{fig:distanceLearningWorldAnalysis} (b)), even though distance education has a long history in contrast to the COVID-19 pandemic.

The speed of the physical classroom closures and the rapid move to online delivery of education allowed little time for planning or reflection on potential risks that could happen to its various stakeholders, such as teachers, students, and parents. Teachers were not fully aware of their obligations and how to connect with students to support their learning. Lack of preparation, lack of tools and techniques for distance education, lack of awareness of the availability of the existing tools and techniques and their effectiveness, lack of interaction and communication with the students, and lack of awareness of students' ongoing problems were some of the major issues they encountered during the COVID-19 pandemic. Students on the other hand who tend to have fewer educational opportunities beyond schools and universities are severely affected by this sudden and unexpected movement. Further, the impact was not limited to students' learning but also to other aspects of their lives, such as student debt, digital exclusion, lack of technology, lack of access to internet-enabled devices or a stable internet connection, long-term educational disengagement, poor nutrition and food insecurity, increased psychological challenges, childcare problems, exploitation, school dropouts and lack of disability services \autocite{drane2020impact,daniel2020education,unescoadverse2020,richmond2020critical,carter2021teacher}. This also put an extra burden on parents, especially those with limited education and resources, as they were expected to facilitate the required learning environment at home. Working parents also find it difficult to work during a school closure due to childcare obligations that result from unexpected school closures. Furthermore, every single step, including planning, developing new tools and techniques, conducting awareness programmes about distance education, and shifting towards distance education, happened online, due to the unexpected massive global shutdown. This situation left the internet as the only medium to support all these educational processes.

According to \textcite{carter2021teacher}, for educators to continue to teach inclusively, they need support to connect with the students through a variety of strategies so that students feel not only connected to their teachers, but to the subject matter being taught. In response to this gray situation and the urgent requirement for a massive transformation from a physical classroom to a virtual learning environment, UNESCO took immediate and timely action by publishing a long list of distance learning solutions \autocite{unesco2020DLsolutions} that can be used to facilitate student learning during the period of school closure. There are many different distance learning solutions. Yet very little is known about the uptake of different solutions or about their effectiveness. Further, getting familiar with all these available distance learning solutions is also not feasible or practical due to limited time availability and the need for urgent responses.

Connectivity, flexibility, and the ability to promote varied interactions are some of the critical aspects educators consider when selecting suitable distance education solutions. According to \textcite{ahn2006utilizing}, the popularity of a product can also greatly influence consumer purchasing decisions as it can indicate the prominence of the product in the market, its usability, and its impact. Further, according to \textcite{willis2020using}, developers also tend to improve their products and services in response to increasing online search. Therefore, the internet represents a great opportunity to learn about the popularity and public attention of a product or service. It is also a way to narrow down the search space and thereby identify the most suitable products and services that meet customer needs.

Google Trends is an open source web analytics tool that allows users to get access to Internet search data that has the ability to provide deep insights into population behavior \autocite{nuti2014use}. As there is still a struggle as to what technologies should be used to facilitate student learning, in this study, we test whether the Google Trends search quiries can be used as a proxy of the popularity and public interest in different distance education solutions. In line with that, this paper makes three fundamental contributions to distance education by exploring three main questions: (1) What is the impact of the COVID-19 pandemic on education? (2) What solutions are in place to meet the needs of distance education during COVID-19 pandemic? (3) Which distance learning solutions have received wide attention and public interest during the COVID-19 pandemic?

We primarily analyzed quantitative digital footprint data on the internet from December 2019 to August 2021. The resulted Google Trend footprint provides a fast first step guide to identifying the most popular distance education tools available for different educational purposes. It also allows the teachers to narrow down the search space and deepen their exploration on prominent distance education solutions to support their online teaching. Furthermore, identifying the most popular tools and techniques will allow everyone to focus on those methods and develop a sustainable learning environment by advancing the existing features.

Quality education is essential to sustainable development and it is a key United Nations (UN) sustainable development goal. Having the most appropriate tool is imperative to the successful completion of a task at hand. The findings of this study provide essential initial guidance to selecting the most appropriate tool for different distance education needs. Therefore, the findings of this study directly contribute to the UN's sustainable development goals of quality education.

The remainder of this paper is organized as follows: Section 2 presents the related work to lay the foundation for the Google trend footprint analysis of different distance learning solutions. Section 3 presents the methodology followed in the study. Section 4 includes results and discussion. Section 5 concludes the article and presents future research directions.

\hypertarget{related-work}{%
\section{Related work}\label{related-work}}

The term distance education is known by a variety of names, such as distance learning, e-Learning, online learning, home study, independent study, external study, correspondence education, off-campus study, open learning, and open education \autocite{moore2011learning}. A few common features found in all these terms are that they all involve some form of knowledge sharing between two parties; learners and educators, different times, different geographical locations and/or varying forms of instructional materials and tools \autocite{moore2011learning}. Further, distance education has different types of interactions, such as learner--content interactions, learner--instructor interactions, learner--learner interactions, and learner--interface interactions that can facilitate learners to construct knowledge \autocite{wallace2003online}. Different tools are required to fulfill all these requirements in a distance learning environment.

Google Trends is a freely accessible website sponsored by Google that analyzes spatio-temporal patterns of search queries in Google Search, across various regions, subject domains, and languages \autocite{carneiro2009google}. Therefore, Google trends provide a quantifiable and valuable measure of emerging public concerns and trending topics together with their geographic distribution \autocite{alicino2015assessing,cook2011assessing}. According to \textcite{jarynowski2020perception}, public concerns on different matters can also take on an epidemic nature, starting with the phase of growing interest, which is known as ``early adoption'', to the phase of general interest, which is known as ``majority'', and eventually lose popularity which is known as ``lagers stage''. Figure \ref{fig:distanceLearningWorldAnalysis} (b) and (c) also confirm this life cycle explanation of Google Trend search patterns.

Google Trend search queries have become a popular data source in monitoring and modeling the dynamics of various application domains, such as disaster management \autocite{kam2019monitoring}, transportation \autocite{willis2020using}, business \autocite{chumnumpan2019understanding}, and health care \autocite{nuti2014use,alicino2015assessing,arora2019google,carneiro2009google,cook2011assessing}, because they can provide data about social phenomena in a more timely manner than than traditional data collection processes \autocite{vaughan2014web}.

According to the study conducted by \textcite{willis2020using} on taxi preferences, online queries can provide a proxy for the quality of taxi services and thereby provide a means to investigate service quality and public concerns about the service provided. In November, 2008, Google launched an internet-based surveillance tool, Google Flu Trends (GFT), that uses aggregated Google search data to provide a near real-time support in predicting influenza outbreaks \autocite{cook2011assessing}. Several recent studies have also been conducted to investigate the relation between Google search trend patterns and the COVID-19 related concerns \autocite{husnayain2020applications,effenberger2020association}. The study conducted by \textcite{husnayain2020applications} focuses on search terms related to the corona virus, hand washing, and face masks to monitor public restlessness toward COVID-19. However, there are only a limited number of research attempts related to education sector using Google Trend search query data \autocite{vaughan2014web}. The study conducted by \textcite{vaughan2014web} have used Google Trend search queries to predict academic fame by testing the correlation between search volume data and university ranking data. \textcite{kansal2021google} have used a text mining approach to identify emerging patterns of words and phrases during COVID-19 pandemic, keeping education prospects as the focus. However, their study was limited to a small number of learning platforms. According to \textcite{moore2011learning}, the design of different types of learning environments can depend on the learning outcomes, target audience, access (physical, virtual and/or both), and/or type of content. Therefore, this study provides a detailed investigation on distance education solutions focusing on different distance learning needs. In contrast to the work in \textcite{kansal2021google}, the work in our paper has a very specific focus and provides a thorough investigation on different distance learning solutions available for various education purposes.

Even though Google Trends search queries provide a massive amount of data, more care should be given when utilizing it as a research tool as it may contain inaccuracies \autocite{carneiro2009google}. According to \textcite{nuti2014use}, the data retrieval process should be transparent. This will increase the trustworthiness of both the results and the generalizability of the findings. Furthermore, \textcite{nuti2014use} believe that researchers should clearly document the rationale and data retrieval process to ensure the reproducibility of results.

\hypertarget{methodology}{%
\section{Methodology}\label{methodology}}

To identify the impact of COVID-19 on education, data on the evolution of school closures and re-openings was obtained from a repository maintained by UNESCO \autocite{unesco2020covid}. Data related to COVID-19 cases was retrieved from the R software \autocite{rsoftware} package \texttt{coronavirus} \autocite{coronavirusr}. The package pulls data from the Johns
Hopkins University Center for Systems Science and Engineering (JHU CCSE) data repository \autocite{dong2020interactive}.

In this work, we also used Google Trend search queries as a measurement proxy to investigate emerging evidence about distance learning and the popularity and public interest in various distance education solutions. We used weekly Google Trends search queries of various distance learning solutions during COVID-19 pandemic. We limited all of our searches to the period from 1 December 2019 to 15 August 2021 to match the period of COVID-19 pandemic. Through this we wanted to capture the volume of interest in the public realm about distance education solutions and how this changes over time.

Google Trends determines categories based on search patterns. According to \textcite{vaughan2014web} more relevant and accurate data can be obtained by limiting the search in Google Trends to a specific category as it helps to reduce noise in the data. In this work, our focus was given to the education category. Google Trends does not provide the absolute search volume of a given search term. Instead, it provides the relative search volume of a particular term, adjusted according to the total searches of the geography and time range it represents. The resulted series scales from 0 to 100, where each data point in the series represents the search interest relative to the highest point in the series for the selected region and time. \autocite{alicino2015assessing,vaughan2014web}. As such, a value of 100 represents the peak popularity of the term.

It is important to decide what term to search for in Google Trends as different terms have different search volumes. In the past, several strategies have been used to select suitable search terms for a given study. Some studies have selected search terms based on intuition and some through brainstorming processes \autocite{vaughan2014web}. In this study, our choice of search term took into account the list of distance learning solutions published by UNESCO \autocite{unesco2020DLsolutions}. Even though, the solutions they have recommended do not carry their explicit endorsement, they tend to have a wide reach, a strong user-base, and evidence of impact \autocite{unesco2020DLsolutions}. For each keyword, the worldwide search queries were performed with the keyword being used as the ``search term''. That allowed us to search for the exact string of text typed by the user. However, according to \textcite{vaughan2014web}, short forms or acronyms had higher search volumes than the corresponding complete names, in general. However, we did not use the acronym in this study, as our focus is on specific tools and techniques available in the market to fulfill distance education needs, and acronyms could be confused with another entity.

We did our analysis under 11 sub-segments namely, digital learning management systems, systems built for use on basic mobile phones, systems with strong offline functionality, Massive Open Online Course (MOOC) platforms, self-directed learning content, mobile reading applications, collaboration platforms that support live-video communication, tools for teachers to create digital learning content, external repositories of distance learning solutions, tools for online proctoring, and resources to provide psychosocial support. The segmentation was mainly motivated by the list of distance learning solutions published by UNESCO in response to the COVID-19 pandemic.

In Google trends, it is possible to search for only up to five queries at a time \autocite{vaughan2014web}. Therefore, under each segmentation, we entered up to five distance learning tools at a time and recorded their relative ranking scores. An iterative pairwise comparison was first used to identify the series that had the highest search volume during the study period. Through this iterative process of relative comparison, we were able to identify the tool with the highest relative ranking score. Then the relative ranking scores for all the other tools in the given segmentation were obtained relative to the previously detected series with the highest ranking scores \autocite{vaughan2014web}.

Both visual and analytical approaches were used to analyse the changes in web search queries at the global level to identify the emerging evidence about the impact of the COVID-19 pandemic on distance education. We also evaluated the relationship between weekly web searches and the number of global cases of COVID-19. Cross correlation analysis and the dynamic time warping algorithm \autocite{giorgino2009computing} were performed to investigate the relationship and similarities between the signatures of COVID-19 related search volume data and distance education related search volume data.

\hypertarget{results-and-discussion}{%
\section{Results and Discussion}\label{results-and-discussion}}

The visual representation of Google Trend search queries of COVID-19 related terms worldwide shows that the highest peak coincided with the start of the pandemic (Figures \ref{fig:distanceLearningWorldAnalysis} (b)). Surprisingly, the search volume index graph for distance education related terms also shows a similar spike at the start of the pandemic (Figures \ref{fig:distanceLearningWorldAnalysis} (c)). This highlights the impact of the COVID-19 pandemic on distance education.

According to \textcite{moore2011learning}, ``distance education'' is the most renowned term used when referencing distance learning. However, both \textcite{benson2002usability} and \textcite{conrad2002deep} introduced ``online learning'' as the most ``recent version,'' ``newer version'' and ``improved version'' of distance learning. The Google Trend search patterns correspond to the term ``online learning'' confirm the above claim as it is the most searched distance education related term throughout the COVID-19 pandemic (Figures \ref{fig:distanceLearningWorldAnalysis} (c)). Despite quantitative differences in terms of relative search volume, the series generated by the two terms ``distance learning'' and ``online teaching'' demonstrate a similar trend during the study period. COVID-19 related web searches declined in the months after the WHO announcement of the COVID-19 pandemic, despite the increase in COVID-19 cases. A similar decline can be observed in distance education-related terms (Figures \ref{fig:distanceLearningWorldAnalysis}). This confirms the life cycle of Google Trend search patterns introduced in \textcite{jarynowski2020perception}. Further, the second peak was observed only in education related terms after the WHO announcement of the COVID-19 pandemic. This could be due to the nearly total closures of schools that happened with the declaration of COVID-19 as a pandemic.

\begin{figure}[h]

{\centering \includegraphics[width=1\textwidth]{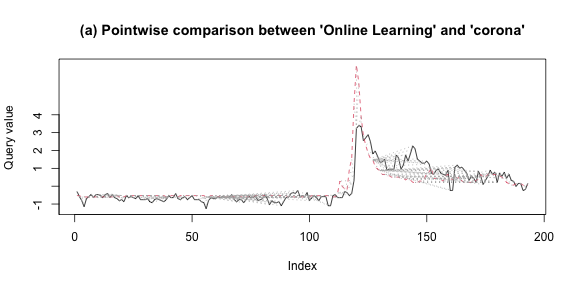} \includegraphics[width=1\textwidth]{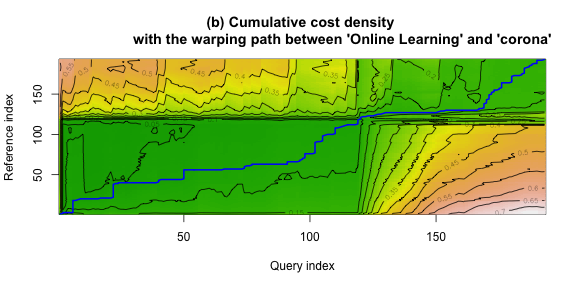} 

}

\caption{Visualization of Dynamic Time Warping Alignments between 'online learning' and 'corona'}\label{fig:dtw1}
\end{figure}

\begin{figure}[h]

{\centering \includegraphics[width=1\textwidth]{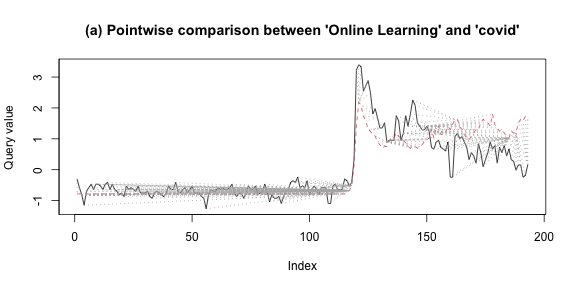} \includegraphics[width=1\textwidth]{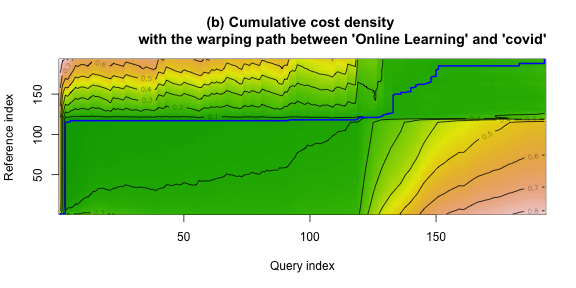} 

}

\caption{Visualization of Dynamic Time Warping Alignments  'online learning' and 'covid'}\label{fig:dtw2}
\end{figure}

A pairwise comparison was performed using the Dynamic Time Warping algorithm. Under this approach, two time series are stretched or compressed locally in order to make one resemble the other as much as possible. The distance between the two series is then computed by summing the distances of individual aligned elements \autocite{giorgino2009computing}. Two pairwise comparisons, ``corona'' with ``online learning'' and ``covid'' with ``online learning'' were conducted to see the impact of COVID-19 on distance education. Both pairwise comparisons show significant areas of overlap and a similar dynamic pattern (Figures \ref{fig:dtw1} (a) and \ref{fig:dtw2} (a)). The normalized distance between the two series in the two comparisons is 0.1326197 and 0.1828299, respectively. This also provides further confirmation of the similarity of the Google Trend search profiles. This highlights the sudden public attention towards both aspects, COVID-19 and distance education, even though distance education has a long history in contrast to COVID-19.

The entirety of the blue warping path flows through green pastures, further confirming the similarity between the search patterns related to COVID-19 and distance education (Figures \ref{fig:dtw1} (b) and \ref{fig:dtw2} (b)). The orange volcanic topography that highlights regions of nonalignment of the series in question if mapped through this region demonstrates a significant difference in online search behaviour before and after the COVID-19 pandemic (Figures \ref{fig:dtw1} (b) and \ref{fig:dtw2} (b)).

The degree of digression of the blue line from the ideal 45-degree straight line indicates a dissimilarity between the two time
series (Figures \ref{fig:dtw1} (b) and \ref{fig:dtw2} (b)). In late 2019 the infection was recognized as a virus which is genetically related to the coronavirus responsible for the Severe Acute Respiratory Syndrome (SARS) outbreak of 2003, which infected over 8,000 people from different countries and resulted in more than 700 deaths worldwide \autocite{ciotti2019covid}. When the virus was known as ``corona'', it was not a global concern and was mostly limited to China. During that time, people did not pay much attention to online learning as conventional classroom learning approach was the most popular teaching learning strategy prior to the COVID-19 pandemic. This is evident from the rough flat line from approximately 120 to 170 on the Query Index axis in Figure \ref{fig:dtw1} (b).

Later, with the spread of the virus worldwide, WHO named it COVID-19 (COrona VIrus Diagnosed in 2019) and then declared it a pandemic on March 11, 2020. With that, a near total closure of schools was observed all around the world (Figures \ref{fig:covidImpactWorld}). With this sudden and unexpected massive shutdown, online learning also became an equally important global concern, similar to the COVID-19 pandemic. This similar behaviour in public attention towards both aspects is confirmed by the blue line that falls on a near 45-degree line from approximately 120 to 150 on the Query Index axis in Figure \ref{fig:dtw2} (b). This also confirms the impact of COVID-19 on distance education and online learning.

\begin{figure}[h]

{\centering \includegraphics[width=1\textwidth]{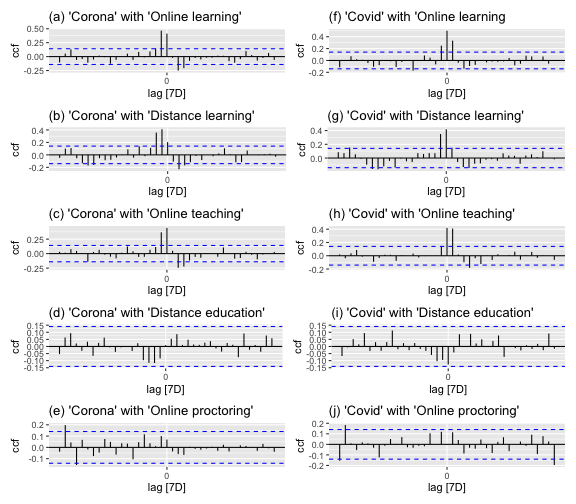} 

}

\caption{Cross-correlation analysis between distance learning related online search terms and COVID-19 relate online search terms during COVID-19 pandemic}\label{fig:ccfAnalysis}
\end{figure}

Next, a cross correlation analysis was performed to estimate the lag time between COVID-19 related terms and the distance education related terms (Figures \ref{fig:ccfAnalysis}). A close relationship was evident between COVID-19 related search patterns and the search patterns for the terms ``Online learning'', ``Distance learning'' and ``Online teaching''. The significant cross-correlation coefficients at lag zero confirm the contemporaneous relationship between the COVID-19 related search terms and distance learning related search terms. The significant cross correlation values at lag one or two also confirm the sudden impact of the COVID-19 pandemic on distance education related public concerns (Figures \ref{fig:ccfAnalysis} (a) - (c) and (f) - (h)). However, no significant synchronization was observed between the COVID-19 related search trends and the search patterns of ``distance education'' or ``online proctoring'' (Figures \ref{fig:ccfAnalysis} (d), (i), (e), (j)).

\begin{figure}[h]

{\centering \includegraphics[width=1\textwidth]{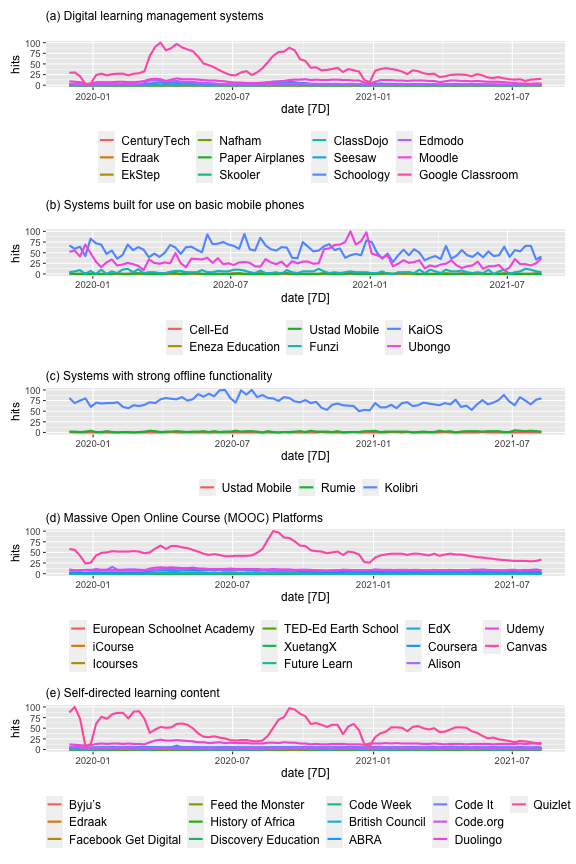} 

}

\caption{Google Trend footprint analysis of distance learning solutions during COVID-19 pandemic}\label{fig:plot1}
\end{figure}

\begin{figure}[h]

{\centering \includegraphics[width=1\textwidth]{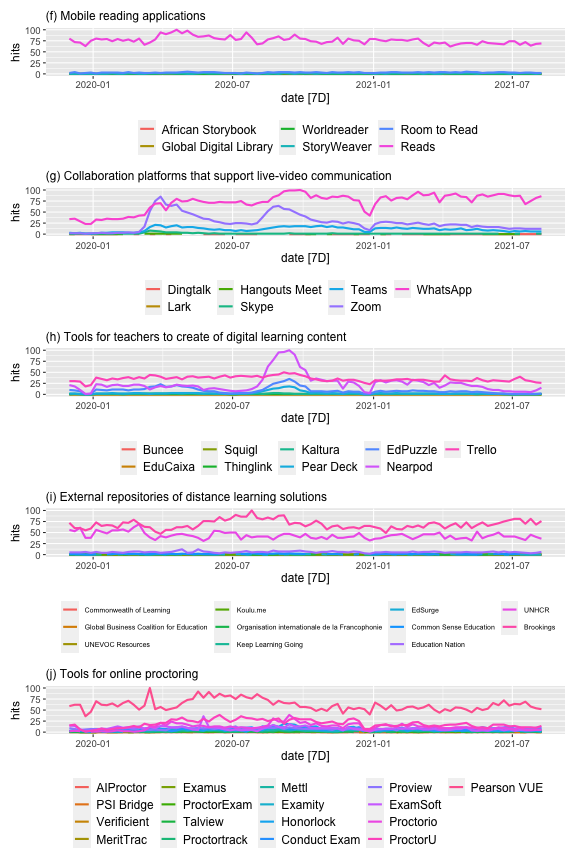} 

}

\caption{Google Trend footprint analysis of distance learning solutions during COVID-19 pandemic}\label{fig:plot2}
\end{figure}

The list of distance education solutions published by UNESCO provides many different distance education solutions for different educational purposes. Figure \ref{fig:plot1} and Figure \ref{fig:plot2} provide a Google trend footprint that can be used as a proxy for measuring global attention towards different distance learning solutions. This provides a fast first step guide to identify the most popular tools available for different educational purposes. In each panel, one or two tools have received significantly more attention than the remaining options in the same category (Figure \ref{fig:plot1} (a) - (e) and Figure \ref{fig:plot2} (f) - (j)). However, the Google Trend series across multiple panels is not comparable as they represent ``relative'' search volume for given terms. They are comparable only within a given panel. The value of 100 is also not comparable across multiple panels and indicates the maximum popularity for the term within each panel.

Under digital learning management systems ``Google Classroom'', which helps teachers and students connect remotely, communicate, and stay-organized, has gotten a wide attention and public interest during the COVID-19 pandemic (Figure \ref{fig:plot1} (a)). This might be due to its easy and fast setup and usage \autocite{sudarsana2019use}. Under systems built for use on basic mobile phones, both `kaiOS' and `Ubongo' have received wide attention during the pandemic (Figure \ref{fig:plot1} (b)).
`KaiOs' has the ability to provide smartphone capabilities to inexpensive, affordable devices and thereby help open portals to learning opportunities. `Ubongo' has also received considerable attention, even though its target groups are school-age children and their parents in Africa. It facilitates localized, multi-platform entertainment-education to African families at a low cost and on a massive scale, and is available in both Kiswahili and English. A significant peak in `Ubongo' series can be observed during the 2020 year end academic break in Africa (Figure \ref{fig:covidImpactContinent}).
Under systems with strong offline functionality, `Kolibri', which is available in more than 20 languages, and supports universal education, has gotten considerable attention during the COVID-19 pandemic (Figure \ref{fig:plot1} (c)) as a tool focuses on facilitating learners living in underserved contexts where the internet is costly, unreliable, or unreachable. These distance learning solutions can provide a huge relief for financially distressed families during the COVID-19 pandemic. Under Massive Open Online Course (MOOC) platforms, `Canvas', which provides free access for teachers to facilitate learning and professional development, has attracted considerable attention in comparison to other options available (Figure \ref{fig:plot1} (d)). `Udemy' has got the second highest attention. However, `Canvas' has received far more attention than `Udemy'. Under self-directed learning content, `Quizlet' has received considerably high attention throughout the pandemic in comparison to the other options available under the same category (Figure \ref{fig:plot1} (e)). It provides learning flashcards, an AI based learning assistant, and games to support student learning in multiple subjects and languages. The second highest option only provides support for language learning. Quizlet's ability to provide support for multiple subjects might be the reason for its ability to outperform its competitors in the market. Under mobile reading applications, `Reads', which provides digital stories with illustrations in multiple languages has consistently been the most popular option (Figure \ref{fig:plot2} (f)). Under collaboration platforms that support live-video communication, both `WhatsApp' and `Zoom' had received a lot of attention at the start of the pandemic in March 2020 (Figure \ref{fig:plot2} (g)). However, `WhatsApp' has outperformed `Zoom' in attracting public attention during the pandemic. In contrast, global attention towards `Zoom' has decreased over time. `Teams' has also gotten considerable attention in comparison to the other remaining options available in the market. During the pandemic, tools for teachers to create digital learning content, such as ``Trello,'' which provides a visual collaboration platform for teachers for coursework planning, classroom organization, and collaboration, received a lot of attention. `Nearpod', which promotes active learning and student engagement through informative and interactive learning activities has also gotten considerable attention during the pandemic (Figure \ref{fig:plot2} (h)). A significant peak can be observed in the `Nearpod' series, during the first academic break. Under external repositories of distance learning solutions, both `Brookings' and `UNHCR' have got significant global attention (Figure \ref{fig:plot2} (i)). `Brookings' provides a large catalogue of learning innovations, though its focus is not limited to distance learning solutions. `UNHCR' also provides an extensive list of distance learning solutions from the United Nations agency for refugees. Student assessment is vital in ensuring the quality of education. However, due to an unexpected and sudden school closure, administering examinations at a distance became a major concern. Under tools for online proctoring, `Pearson VUE' which provides computer based testing services in
over 150 countries worldwide with more than 5,000 authorized test centers has attracted considerably high public attention during the pandemic (Figure \ref{fig:plot2} (j)).

At the start of the pandemic in March 2020, a significant peak of public attention can be observed in both digital learning management systems (Figure \ref{fig:plot1} (a)) and collaboration platforms that support live-video communication (Figure \ref{fig:plot2} (g)). These peaks align with the near-total closures of schools due to the COVID-19 pandemic (Figure \ref{fig:covidImpactWorld}). The major concerns that arise with the unexpected, sudden movement from a physical classroom to a virtual learning environment, such as a lack of experience in connecting and sharing learning materials with students could be the reason for this notable public attention towards these two aspects. Further, a significant peak can be observed in digital learning management systems (Figure \ref{fig:plot1} (a)), Massive Open Online Course (MOOC) Platforms (Figure \ref{fig:plot1} (d)), self-directed learning content (Figure \ref{fig:plot1} (e)), collaboration platforms that support live-video communication (Figure \ref{fig:plot2} (g)) and tools for teachers to create digital learning content (Figure \ref{fig:plot2} (h)), immediately after the academic break in August, 2020 (Figure \ref{fig:covidImpactWorld}). The inability to reopen schools fully due to the increasing COVID-19 cases worldwide and the need for urgent responses to connect with students to somehow facilitate their learning could be the reasons for this considerable public attention around August, 2020.

\begin{figure}[h]

{\centering \includegraphics[width=1\textwidth]{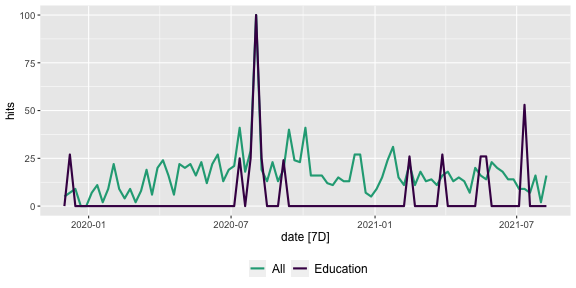} 

}

\caption{Goolge search trend of the term psychosocial support under All categories and Education category during COVID-19 pandemic}\label{fig:psychosocialSupport}
\end{figure}

Good psychosocial support is also vital to improving both students' and teachers' mental health and well-being during a pandemic. This is evident from the two sudden spikes in the search volume index graphs for the term `Psychosocial support' ( Figure \ref{fig:psychosocialSupport} ) which correspond well with the academic breaks in August in both years, 2020 and 2021, under `Education' category. This highlights the urgent need for attention and solutions to these aspects. This type of analysis based on secondary data is vital as these aspects can be easily overlooked due to the lack of connection and communication between students and educators during school closures.

\hypertarget{conclusions-and-further-work}{%
\section{Conclusions and Further work}\label{conclusions-and-further-work}}

In this study, we investigated whether the Google Trend search queries can be used as a proxy of the popularity and public interest in different distance education solutions. This study is the first detailed analysis of web search behaviour related to distance learning solutions during the COVID-19 pandemic, both in quantitative and qualitative terms. A few previous published studies \autocite{vaughan2014web,kansal2021google} looked at Google trend search queries related to education, but only with a limited number of search terms or with different focuses.

Our findings suggest that web search interest in distance education at the global level was strongly influenced by the COVID-19 pandemic. Google Trend search queries showed a strong cross correlation between COVID-19 related search terms and distance education related search terms. The high popularity and increasing public attention towards different distance learning solutions during the COVID-19 pandemic was also investigated through a detailed Google Trend footprint analysis. The findings will help educators to narrow down their search space and thereby select prominent options available in the market for different educational purposes. According to \textcite{wallace2003online}, the learning tools can have a significant impact on how students engage in online learning. The finding will also be helpful for developers to identify their competitors in the market. The high costs involved with different distance learning solutions, lack of financial support/funding, and lack of understanding about the existing distance learning solutions are some of the major issues educational institutions face with the COVID-19 pandemic. Therefore, the findings of this study offer important insights to various stakeholders in education to make better decisions with less time and effort.

Google Trends only captures the search behaviour of a sub-population with Internet access, even though it is the most common worldwide search engine. Furthermore, our study focused on global attention toward distance learning solutions during the COVID-19 pandemic. Therefore, care should be given when generalizing these findings to different regions or different time periods, as different regions can have different market orientation and seasonal requirements. This can lead to variations in the search output and in turn the study findings for different regions and time frames. Furthermore, Google Trends data is not available as absolute values but only in the form of relative volume. Therefore, the scores range from 0 to 100, which does not allow the researchers to make meaningful comparisons across different education needs. Further investigation is needed at national and regional level to identify the specific measures that can be taken to ensure the quality of distance education at national and regional level.

\hypertarget{supplementary-materials}{%
\section*{Supplementary Materials}\label{supplementary-materials}}
\addcontentsline{toc}{section}{Supplementary Materials}

\textbf{Data and scripts}: Datasets and R code to reproduce all figures in this article (\url{https://github.com/pridiltal/Covid19_Online_Teaching_paper}).

\hypertarget{acknowledgment}{%
\section*{Acknowledgment}\label{acknowledgment}}
\addcontentsline{toc}{section}{Acknowledgment}

This work was carried out with the aid of a grant from UNESCO and the International Development Research Centre, Ottawa, Canada. The views expressed herein do not necessarily represent those of UNESCO, IDRC or its Board of Governors.

\printbibliography

\end{document}